# Magnetic design study of coil-dominated superconducting quadrupole magnets based on racetrack coils

Chuang Shen, Yingshun Zhu*, Fusan Chen

*Abstract*—Several coil structures have been used in accelerator superconducting quadrupole magnets, and cos2$\theta$ quadrupole magnets are the most mature in theoretical research and engineering applications. However, the cos2$\theta$ quadrupole magnet has a complicated coil structure, especially at the end of the coil, which makes it difficult to apply strain-sensitive high-temperature superconductors. Racetrack quadrupole magnets are friendly to high-temperature superconductors. Field strength of iron-dominated racetrack magnets is limited by the magnetic saturation of the iron poles. Therefore, coil-dominated racetrack quadrupole magnets with simple geometry have become the focus of our research. In this paper, analytical expressions of the magnetic field harmonics related to racetrack quadrupole coil parameters are obtained. These expressions are used to find the solution of coil geometry parameters with field harmonics on the order of $10^{-4}$. Then, examples are given to build ideal quadrupole model and verify the theoretical formulas. Next, the design and optimization of example racetrack quadrupole magnets are completed in ROXIE. Finally, the advantages and disadvantages of the racetrack coils and the cos2$\theta$ coils are compared and discussed.

*Index Terms*—Superconducting quadrupole magnet, racetrack coil, HTS YBCO, LTS NbTi, magnetic design, field harmonics, coil structure.

## I. INTRODUCTION

With the development of superconducting magnet technology, coil structures including cos2$\theta$ coils, serpentine coils and canted cosine theta (CCT) coils have been greatly developed and applied in superconducting quadrupole magnets. Cos2$\theta$ coils are one of the most commonly used coil structures for superconducting quadrupole magnets. After decades of development, cos2$\theta$ superconducting magnets are quite mature in theoretical guidance and engineering applications [1]-[2]. The most advanced cos2$\theta$ quadrupole magnet is the superconducting magnet MQXF using $Nb_3Sn$ superconductors for the HL-LHC upgrade [3]-[4].

The current distribution in a CCT multipole magnet can be written starting from the equation of an azimuthal modulated helix [5]. The superconducting quadrupole magnets with CCT structure also have complete theoretical support and the superconducting quadrupole coils with CCT structure were used in the preliminary study of the superconducting quadrupole magnets in the interaction region of FCC-ee [6]. Serpentine winding, an innovation developed at BNL for direct winding superconducting magnets, allows winding a coil layer of arbitrary multipolarity in one continuous winding process and greatly simplifies magnet design and production compared to the planar patterns used before. Serpentine windings were used for the BEPC-II Upgrade and JPARC magnets and are proposed to make compact final focus magnets for the ILC [7]-[8].

With the deepening of particle physics research, the beam energy of particle accelerators is constantly increasing. As the beam energy is further increased, the higher gradient superconducting quadrupole magnets are required to focus the beam. The higher gradient means a large current carrying capacity of the superconductors in high magnetic field. High-temperature superconductors (HTS) have the potential to generate a magnetic field beyond the level obtained with low-temperature superconductors (LTS) [9]-[10]. However, High-temperature superconductors are more strain-sensitive than low-temperature superconductors and are not suitable for cos2$\theta$ coils with complex spatial geometries. Therefore, the racetrack quadrupole coils with simple geometry are a better choice for high-temperature superconductors.

Iron dominated and coil dominated are the two basic types of superconducting magnets [11]-[12]. The racetrack quadrupole magnets of iron dominated have low magnetic field intensity due to the saturation phenomenon of their iron poles [13]-[14], which cannot meet the requirements of the large particle accelerator. From this point of view, coil dominated racetrack quadrupole magnet is a suitable scheme to achieve ultra-high magnetic field strength.

This work was supported in part by the National Natural Science Foundation of China under Contract 11875272 and in part by the Key Laboratory of Particle Acceleration Physics and Technology, Institute of High Energy Physics, Chinese Academy of Sciences. (*Corresponding author: Yingshun Zhu.)

Chuang Shen, Yingshun Zhu, Fusan Chen are with the Institute of High Energy Physics, Chinese Academy of Sciences, Beijing 100049, China, and also with the University of Chinese Academy of Science, Beijing 100049, China(e-mail: shenchuang@ihep.ac.cn; yszhu@ihep.ac.cn;chenfs@ihep.ac.cn).

In recent years, some research on coil-dominated racetrack quadrupole magnets has been done [15]-[17], but the research is insufficient and needs to be further improved. Firstly, the design of the racetrack quadrupole magnets lacks theoretical guidance. Secondly, additional return coil blocks are added to form racetrack coils, which increases the size of magnet cross section, and the current distribution patterns are complex. Finally, the high-order field harmonics are large and the field quality is difficult to meet the needs of practical engineering.

In this paper, theoretical study of coil-dominated superconducting quadrupole magnets based on racetrack coils is introduced. First of all, analytical formulas of racetrack superconducting quadrupole coils are derived, including high-order field harmonics, field gradient, etc. Then, using the derived formulas, an ideal quadrupole magnet model based on racetrack coils is established and the corresponding coil parameters are solved, with the field harmonics on the order of $10^{-4}$. Next, the racetrack quadrupole magnets are designed in the electromagnetic software ROXIE. Finally, under the same design requirements, the $\cos 2\theta$ and racetrack quadrupole coils are respectively established and the important parameters are compared and discussed.

## II. THEORETICAL STUDY OF RACETRACK QUADRUPOLE COIL

According to the complex formalism, a line carrying a current $I$ in the position $z_0 = x_0 + iy_0$ generates a magnetic field $B(z) = B_y(z) + iB_x(z)$ in the position $z = x + iy$ that reads

$$B(z) = \frac{I\mu_0}{2\pi(z - z_0)} \tag{1}$$

One can expand the series as

$$B(z) = -\frac{I\mu_0}{2\pi z_0} \sum_{n=1}^{\infty} \left(\frac{z}{z_0}\right)^{n-1}$$

$$= -\frac{I\mu_0}{2\pi z_0} \sum_{n=1}^{\infty} \left(\frac{R_{ref}}{z_0}\right)^{n-1} \left(\frac{z}{R_{ref}}\right)^{n-1} \tag{2}$$

Where $R_{ref}$ is the reference radius, usually chosen as 2/3 of the aperture radius, and $\mu_0$ is the vacuum permeability. The multipolar expansion of the magnetic field according to the European notation (n=1 being the dipole) reads

$$B(z) = \sum_{n=1}^{\infty} C_n \left(\frac{z}{R_{ref}}\right)^{n-1}$$

$$= \sum_{n=1}^{\infty} (B_n + iA_n) \left(\frac{z}{R_{ref}}\right)^{n-1} \tag{3}$$

From the above two expressions, the following equation can be obtained

$$B_n + iA_n = -\frac{I\mu_0}{2\pi z_0} \left(\frac{R_{ref}}{z_0}\right)^{n-1}$$

$$= -\frac{I\mu_0}{2\pi} \frac{R_{ref}^{n-1}}{z_0^n} \tag{4}$$

According to $z_0^n = r_0^n e^{in\theta_0} = r_0^n (\cos n\theta_0 + i \sin n\theta_0)$.

$$B_n + iA_n = -\frac{I\mu_0}{2\pi} \frac{R_{ref}^{n-1}}{r_0^n} \cos n\theta_0 + i \frac{I\mu_0}{2\pi} \frac{R_{ref}^{n-1}}{r_0^n} \sin n\theta_0 \tag{5}$$

So the multipole field in the aperture generated by a line current $I$ at position $z_0 = x_0 + iy_0 = r_0 e^{i\theta_0}$ is

$$B_n = -\frac{I\mu_0}{2\pi} \frac{R_{ref}^{n-1}}{r_0^n} \cos n\theta_0 \tag{6}$$

$$A_n = \frac{I\mu_0}{2\pi} \frac{R_{ref}^{n-1}}{r_0^n} \sin n\theta_0 \tag{7}$$

Normal quadrupole magnets and skew quadrupole magnets are two basic quadrupole magnets. Normal quadrupole magnet can be obtained by rotating skew quadrupole magnet 45 degrees. The multipole field in the aperture generated by a rotated line current $I'$ at position $z_0' = z_0 \cdot e^{-i\frac{\pi}{4}} = r_0 e^{i(\theta_0 - \frac{\pi}{4})}$ is

$$B_n' = -\frac{I\mu_0}{2\pi} \frac{R_{ref}^{n-1}}{r_0^n} \cos n(\theta_0 - \frac{\pi}{4}) \tag{8}$$

$$A_n' = \frac{I\mu_0}{2\pi} \frac{R_{ref}^{n-1}}{r_0^n} \sin n(\theta_0 - \frac{\pi}{4}) \tag{9}$$

For the high-order field harmonics of quadrupole magnetic field, the $n$ in equation (8) equals 4N-2, N=1, 2, 3…. The equation (8) can be written

$$B_n' = -\frac{I\mu_0}{2\pi} \frac{R_{ref}^{n-1}}{r_0^n} \cos\left((2N-1)\frac{\pi}{2} - (4N-2)\theta_0\right) \tag{10}$$

When $N$ is odd, namely $n = 2, 10, 18\ldots$

$$\begin{aligned}B'_n &= -\frac{I\mu_0}{2\pi}\frac{R_{ref}^{n-1}}{r_0^n}\sin\bigl((4N-2)\theta_0\bigr)\\ &= -\frac{I\mu_0}{2\pi}\frac{R_{ref}^{n-1}}{r_0^n}\sin(n\theta_0)\\ &= -A_n\end{aligned} \quad (11)$$

When N is even, namely $n = 6, 14, 22\ldots$

$$\begin{aligned}B'_n &= \frac{I\mu_0}{2\pi}\frac{R_{ref}^{n-1}}{r_0^n}\sin\bigl((4N-2)\theta_0\bigr)\\ &= \frac{I\mu_0}{2\pi}\frac{R_{ref}^{n-1}}{r_0^n}\sin(n\theta_0)\\ &= A_n\end{aligned} \quad (12)$$

According to equation (11) and equation (12), we can obtain

$$b'_n = \frac{B'_n}{B'_2} = \begin{cases}\dfrac{-A_n}{-A_2} = a_n & \text{if } N \text{ is odd}, n = 4N-2 = 2, 10, 18\cdots\\ \dfrac{A_n}{-A_2} = -a_n & \text{if } N \text{ is even}, n = 4N-2 = 6, 14, 22\cdots\end{cases} \quad (13)$$

To sum up, among the high-order harmonics $a_6$, $a_{10}$ and $a_{14}$ of the quadrupole magnetic field, a negative sign needs to be added to the harmonics $a_6$ and $a_{14}$ in the transformation from the skew quadrupole magnet to the normal quadrupole magnet.

Though people are usually interested in normal quadrupole magnets, to facilitate the integral calculation process in the Cartesian coordinate system, in the following sections skew quadrupole magnet is taken as our research object. A normal quadrupole magnet can be obtained when skew quadrupole magnet is rotated by 45 degrees. Correspondingly, the multipole field also changes from skew field $A_n$ to normal field $B_n$, and such a change only affects the positive or negative signs of the field harmonics.

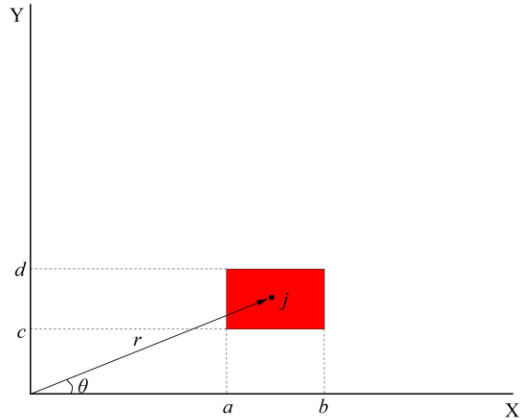

Fig. 1. Parameters definition of a rectangular current block.

The parameters definition of a rectangular current block is shown in Fig. 1. To derive the total field generated by the current block, firstly the expression of multipole field in the polar coordinate system is transformed into the Cartesian coordinate system. The transformation formula of the two coordinate systems reads

$$\begin{cases}r = \sqrt{x^2 + y^2}\\ \theta = \arctan\left(\dfrac{y}{x}\right)\end{cases} \quad (14)$$

From equations (7) and (14), the multipole field generated by differential unit of the rectangular block is

$$dA_{n(block-abcd)} = \frac{\mu_0 j}{2\pi}\frac{R_{ref}^{n-1}}{(x^2+y^2)^{\frac{n}{2}}}\sin n\left(\arctan\left(\frac{y}{x}\right)\right)dydx \quad (15)$$

Where $j$ represents current density of the coil section. Multipole field in the aperture generated by a rectangular current block shown in Fig. 1 can be calculated by

$$A_{n(block-abcd)} = \int_a^b\int_c^d \frac{\mu_0 j}{2\pi}\frac{R_{ref}^{n-1}}{(x^2+y^2)^{\frac{n}{2}}}\sin n\left(\arctan\left(\frac{y}{x}\right)\right)dydx \quad (16)$$

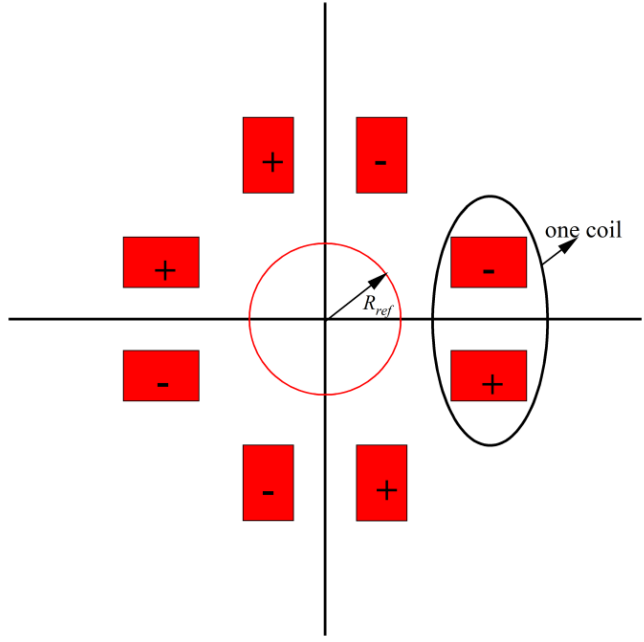

Fig. 2. Cross-section of skew racetrack quadrupole coils.

Full cross section of a skew quadrupole magnet with racetrack coils is shown in Fig. 2. There are a total of eight rectangular current blocks, namely four racetrack coils. According to the symmetry of quadrupole magnet, the multipole field in the aperture generated by the skew quadrupole coils reads

$$A_n = 8 \int_a^b \int_c^d \frac{\mu_0 j}{2\pi} \frac{R_{ref}^{n-1}}{(x^2+y^2)^{\frac{n}{2}}} \sin n\left(arctan\left(\frac{y}{x}\right)\right) \mathrm{d}y \mathrm{d}x \tag{17}$$

### III. SOLUTION OF RACETRACK QUADRUPOLE COILS USING THEORETICAL EXPRESSIONS

#### A. Quadrupole coils with one rectangular block

According to the equation (17), each order multipole field $A_n$ of racetrack quadrupole coils can be obtained. For example, we can get the analytic expressions of quadrupole field component and field gradient

$$A_2 = \frac{2\mu_0 j R_{ref}}{\pi} \ln \frac{(a^2+d^2)(b^2+c^2)}{(a^2+c^2)(b^2+d^2)} \tag{18}$$

$$G = \frac{A_2}{R_{ref}} = \frac{2\mu_0 j}{\pi} \ln \frac{(a^2+d^2)(b^2+c^2)}{(a^2+c^2)(b^2+d^2)} \tag{19}$$

The relative field harmonics normalized to the main quadrupole field component can be defined by $a_n = A_n/A_2$. Then the expressions of systematic field harmonics $a_6$, $a_{10}$, and $a_{14}$ can all be derived. These expressions contain only four parameters, namely $a$, $b$, $c$, and $d$, and are listed in the appendix because they are too long.

In the following, we use an application example to verify the correctness of the formulas and give the specific magnitude of magnetic field harmonics. The reference radius $R_{ref}$ is set to 9.8 mm.

Parameter $a$ represents the X-axis coordinate value corresponding to the left side of the rectangular coil section, which is directly related to the aperture of the racetrack superconducting quadrupole magnet. Therefore, after reserving the space required by the coil support structure, parameter $a$ is the first variable determined in the design process. Considering the aperture of our example quadrupole coil, parameter $a$ is set to 0.02 m. Then we put it into the theoretical formulas to study the relationship between the high-order field harmonics and each coil parameter. Among high-order field harmonics of superconducting quadrupole magnets, the systematic harmonics $a_6$, $a_{10}$ and $a_{14}$ are our main research objects. Using the analytic expression in the appendix, the field harmonics $a_6$ as a function of the coil parameters is shown in Fig. 3.

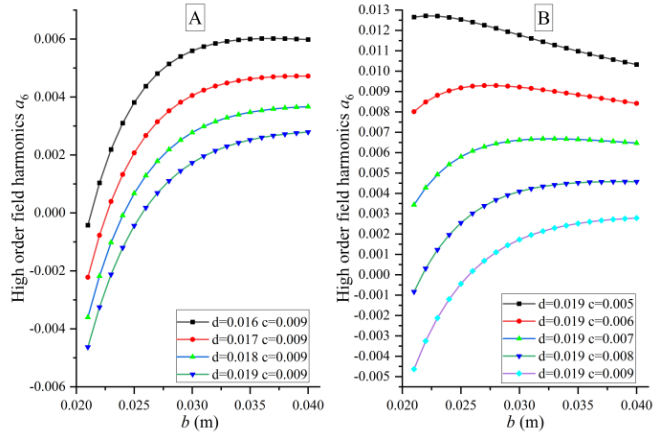

Fig. 3. High-order harmonics $a_6$ under different parameters $c$ and $d$.

From the analysis of the high-order harmonics of $a_6$, $a_{10}$, and $a_{14}$, it is not difficult to find that the $a_{14}$ is relatively small. So $a_6$ and $a_{10}$ are the keys to our research. We need to find a two-dimensional rectangular block that meets the requirements of high-order field harmonics on the order of $10^{-4}$. From Fig. 3, we can get a group of parameters with a relatively small value of $a_6$. Then $a_6$ and $a_{10}$ as a function of coil parameters are shown in Fig. 4.

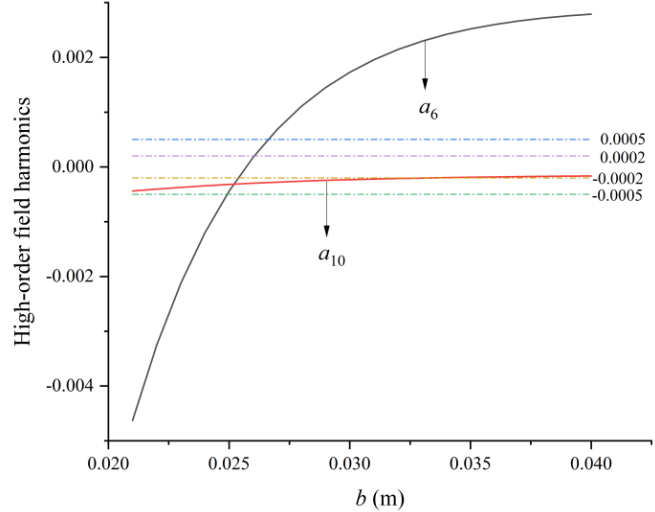

Fig. 4. High-order harmonics with parameters $a$=0.02, $c$=0.009, $d$=0.019.

It can be seen from Fig. 4 that the parameters of a rectangular cross-section can be found at the harmonic requirement of less than $5×10^{-4}$. When field harmonic requirements are further raised to be smaller than $2×10^{-4}$, one rectangular block of the quadrupole coil cannot meet the requirements.

*B. Quadrupole coils with two rectangular blocks*

When the physical requirements of high-order field harmonics are stricter, we consider the superconducting quadrupole coil composed of two or more rectangular blocks.

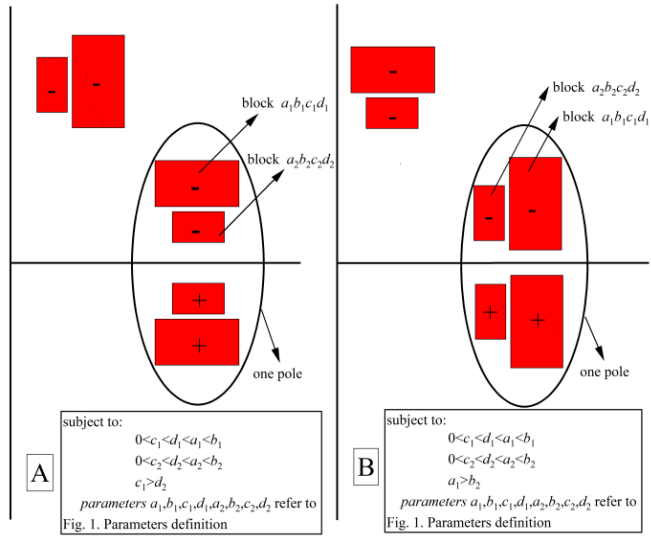

Fig. 5. Cross-section of the two racetrack quadrupole coils.

Two kinds of racetrack quadrupole coils are shown in Fig. 5. For the range of 0 to 45 degrees in the quadrant of 2-D Cartesian coordinate system, the two rectangular blocks $a_1b_1c_1d_1$ and $a_2b_2c_2d_2$ have the following constraints: $0 < c_1 < d_1 < a_1 < b_1$ and $0 < c_2 < d_2 < a_2 < b_2$. For the two different coil structures, to ensure that the two rectangular blocks do not overlap, we add the following constraints: $(c_1 > d_2)$ and $(a_1 > b_2)$.

According to equation (18) and the analytic expressions in the appendix, the high-order harmonics of racetrack quadrupole coils consisting of two blocks are obtained.

$$a_{n-two\ blocks} = \frac{A_{n-two\ blocks}}{A_{2-two\ blocks}} = \frac{A_{n-block\ a_1b_1c_1d_1} + A_{n-block\ a_2b_2c_2d_2}}{A_{2-block\ a_1b_1c_1d_1} + A_{2-block\ a_2b_2c_2d_2}} \quad (20)$$

According to our design requirements, we can get the following constraints.

$$\begin{cases} \left| a_6 = \frac{A_{6-block\ a_1b_1c_1d_1} + A_{6-block\ a_2b_2c_2d_2}}{A_{2-block\ a_1b_1c_1d_1} + A_{2-block\ a_2b_2c_2d_2}} \right| \leq 1 \times 10^{-4} \\ \left| a_{10} = \frac{A_{10-block\ a_1b_1c_1d_1} + A_{10-block\ a_2b_2c_2d_2}}{A_{2-block\ a_1b_1c_1d_1} + A_{2-block\ a_2b_2c_2d_2}} \right| \leq 1 \times 10^{-4} \\ \left| a_{14} = \frac{A_{14-block\ a_1b_1c_1d_1} + A_{14-block\ a_2b_2c_2d_2}}{A_{2-block\ a_1b_1c_1d_1} + A_{2-block\ a_2b_2c_2d_2}} \right| \leq 1 \times 10^{-4} \end{cases} \quad (21)$$

This is a complex set of inequalities. Considering the factors in the actual design process, such as the inner diameter and the bending radius of the coil, we will get the specific range of parameters. Then, considering the characteristics of superconductors, such as the width and thickness of superconductors, we set the step size of parameter search. Finally, we find the parameters satisfying the high-order harmonics requirements by global search. Of course, there are many solutions that meet all our requirements. We select and list two typical solutions in TABLE Ⅰ.

TABLE I
COMPARISON OF THE ANALYTICAL RESULTS AND THE FEM RESULTS
(IN UNITS OF $10^{-4}$)

|  | Analytical | FEM(ROXIE) | Analytical | FEM(ROXIE) |
| --- | --- | --- | --- | --- |
| Block | $a_1$=0.02<br>$b_1$=0.026<br>$c_1$=0.012<br>$d_1$=0.019 | $a_2$=0.02<br>$b_2$=0.026<br>$c_2$=0.006<br>$d_2$=0.007 | $a_1$=0.0234<br>$b_1$=0.026<br>$c_1$=0.008<br>$d_1$=0.018 | $a_2$=0.02<br>$b_2$=0.0226<br>$c_2$=0.015<br>$d_2$=0.019 |
| Gradient T/m | 93.490 | 93.4899 | 92.194 | 92.1938 |
| $a_6$(units) | 0.88194 | 0.88188 | -0.05648 | -0.05422 |
| $a_{10}$(units) | -0.56287 | -0.56286 | -0.97084 | -0.9706 |
| $a_{14}$(units) | -0.016947 | -0.01695 | -0.007166 | -0.00717 |

According to the solution results of analytical method, quadrupole model with the same coil parameters is established in software ROXIE [20], and the comparison between the results of analytical method and Finite Element Method is also shown in Table I. It can be seen that the results of analytical method and FEM are consistent with each other.

## IV. MAGNET DESIGN OF EXAMPLE RACETRACK QUADRUPOLE

The above content completes the theoretical study of the racetrack quadrupole coil and the geometric parameters of the ideal racetrack coil that meet the physical requirements are obtained. Next, we will perform electromagnetic design of the example racetrack quadrupole magnet using the professional software ROXIE. The design requirements of the example superconducting racetrack quadrupole magnet (normal magnet) are listed in TABLE Ⅱ.

TABLE Ⅱ
DESIGN REQUIREMENTS OF THE SUPERCONDUCTING QUADRUPOLE MAGNET

| Parameters | Value |
| --- | --- |
| The inner radius of the quadrupole coils | 20 mm |
| Field gradient | 136 T/m |
| Reference radius | 9.8 mm |
| High-order field harmonics | ≤3×10$^{-4}$ |

We already obtained parameters $a_1, b_1, c_1, d_1, a_2, b_2, c_2, d_2$ of the skew racetrack quadrupole coils in Table Ⅰ. According to the conversion of normal quadrupole and skew quadrupole, the vertex coordinates of two blocks in skew quadrupole are rotated by 45 degrees to become the vertex coordinates of two blocks in normal quadrupole. Then, we can create the normal quadrupole coils model in ROXIE using those coordinate points directly.

In the ideal model study, we obtained two kinds of racetrack coil structures, namely single-layer quadrupole coil structure and double-layer quadrupole coil structure. For the single-layer normal racetrack quadrupole coils created in ROXIE, high-temperature superconductor YBCO is used in our magnet design. The width of the YBCO cable is 6 mm and the thickness is 0.2 mm. As shown in Fig. 6, each pole has 40 turns, and the peak field in coils is 3.077 T. The design value of the operating current is 1250 A. The calculated field harmonics are all within 1×10$^{-4}$, and the detailed parameters is shown in Table Ⅲ.

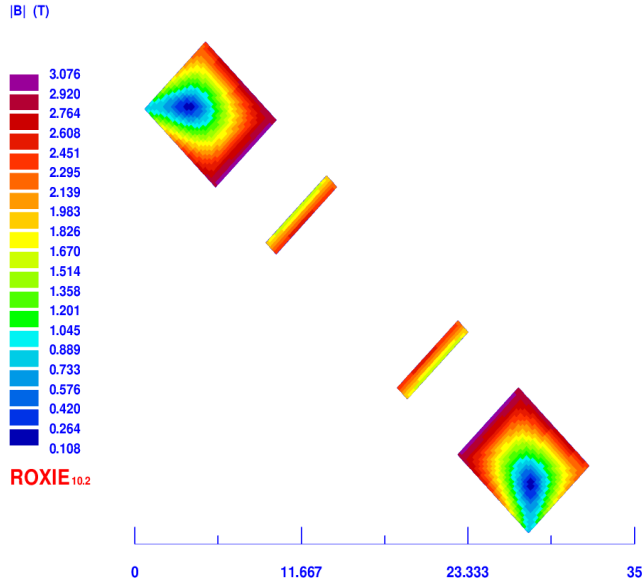

Fig. 6. Simulation result of single-layer racetrack quadrupole coils in ROXIE.

TABLE Ⅲ
2D SIMULATION RESULTS OF THE SINGLE-LAYER QUADRUPOLE COILS
(IN UNITS OF 10$^{-4}$)

| Parameters | Value |
| --- | --- |
| Current | 1250 A |
| Current density in strands | 1041.667 A/mm$^2$ |
| Gradient | 93.4899 T/m |
| Peak field in coils | 3.077 T |
| Cross-sectional area of one coil | 48 mm$^2$ |
| $b_6$ (units) | -0.88188 |
| $b_{10}$ (units) | -0.56286 |
| $b_{14}$ (units) | 0.01695 |

For the double-layer normal racetrack quadrupole coils, the width of the YBCO cable is 2.6 mm and the thickness is 0.2 mm. The cross-sectional layout of the 2D coil is obtained in Fig. 7. Each pole has 70 turns, and the operating current is 745 A. The high-order field harmonics are all within $1\times10^{-4}$.

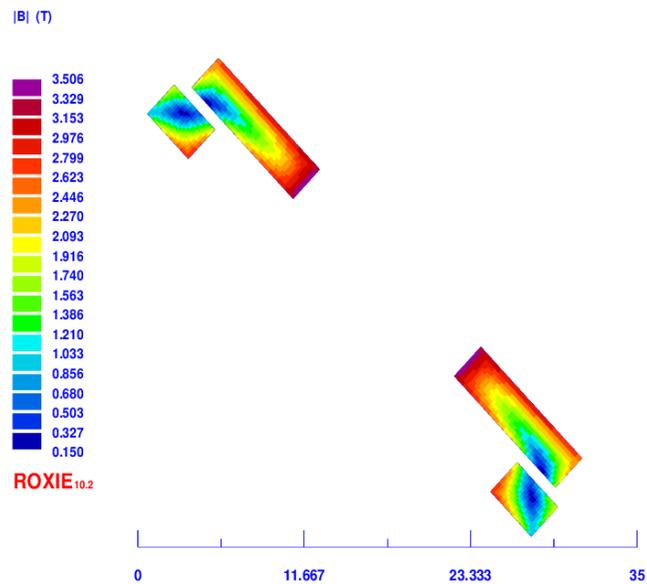

Fig. 7. Simulation result of double-layer racetrack quadrupole coils in ROXIE.

TABLE IV
2D SIMULATION RESULTS OF THE DOUBLE-LAYER QUADRUPOLE COILS
(IN UNITS OF $10^{-4}$)

| Parameters | Value |
| --- | --- |
| Current | 745 A |
| Current density in strands | 1432.692 A/mm$^2$ |
| Gradient | 92.1938 T/m |
| Peak field in coils | 3.506 T |
| Cross-sectional area of coils | 36.4 mm$^2$ |
| $b_6$ (units) | 0.05422 |
| $b_{10}$ (units) | -0.9706 |
| $b_{14}$ (units) | 0.00717 |

From Table Ⅰ, Table Ⅲ and Table Ⅳ, we can see that the amplitude of the magnetic field harmonics is the same in normal or skew quadrupole coils, but the signs of some harmonics have changed.

To enhance the magnetic field gradient, we add an iron shell outside of the coil. As shown in Fig. 8, the inner edge of the iron is a regular octagon. The radius of the inner tangent circle of the octagon is 32 mm, and the outer radius of the iron shell is 54 mm. The design value of the operating current is 1250 A. The calculated high-order harmonics are listed in Table Ⅴ.

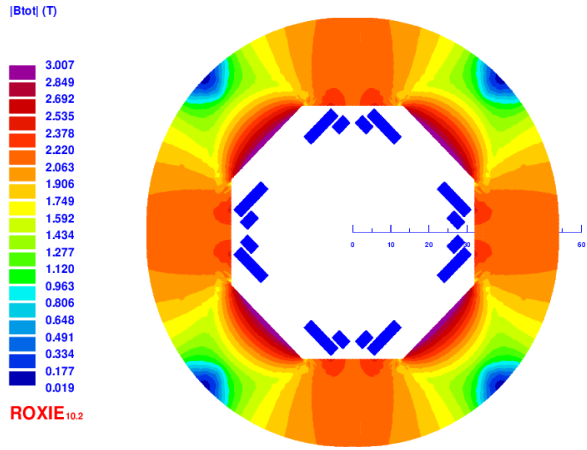

Fig. 8. Simulation result of double-layer racetrack quadrupole coils with iron.

TABLE V
2D SIMULATION RESULTS OF THE DOUBLE-LAYER COIL WITH IRON
(IN UNITS OF $10^{-4}$)

| Parameters | Value |
| --- | --- |
| Current | 745 A |
| Current density in strands | 1432.692 A/mm$^2$ |
| Gradient | 136.2452 T/m |
| Peak field in coils | 4.574 T |
| $b_6$ (units) | -1.60010 |
| $b_{10}$ (units) | -0.63153 |
| $b_{14}$ (units) | 0.00481 |

Compared to the values in Table IV, when iron is added, the field harmonics increase slightly, but are still within $2\times10^{-4}$. Now the field gradient meets the design requirement.

To reflect the structure of the racetrack magnet intuitively, the three-dimensional structure of the magnet is constructed, as shown in Fig. 9. The integrated high-order field harmonics are all smaller than $2\times10^{-4}$. Racetrack coils are simpler than complex cos$2\theta$ coils. Especially at the end, the racetrack coil turns more gently. So the winding of racetrack coils and the manufacturing of coil parts are more convenient.

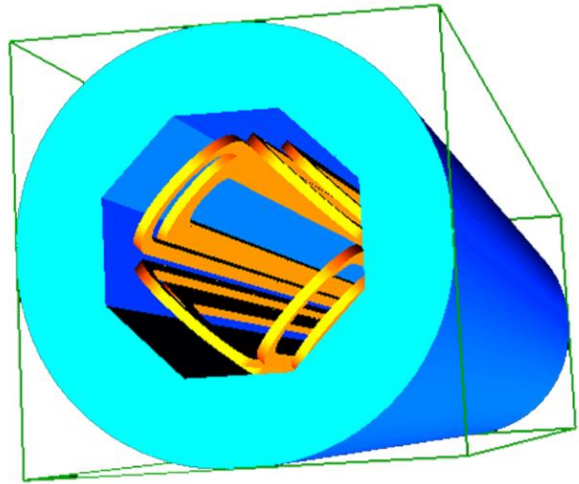

Fig. 9. 3D magnet model.

The above-mentioned two kinds of racetrack quadrupole magnets are theoretically feasible. In our study, only two rectangular blocks are used to adjust the high-order field harmonics, and the systematic field harmonics reach the precision of $1\times10^{-4}$. In specific engineering applications, more layers of rectangular blocks may be needed to meet the design requirements, mainly to increase the field gradient. Furthermore, other HTS superconductor, such as Bi-2212, can also be used in superconducting quadrupole magnets based on racetrack coils.

## V. COMPARISON OF $\cos 2\theta$ QUADRUPOLE COIL AND RACETRACK QUADRUPOLE COIL

Under the same physical design requirements listed in TABLE II, the design of the superconducting quadrupole magnet based on $\cos 2\theta$ coil was performed. The inner diameter of the coil is required to be 20 mm, and the field gradient is 136 T/m. The inner diameter and the outer diameter of the iron are 32 mm and 54 mm, respectively.

In the design of the $\cos 2\theta$ quadrupole magnet, the Rutherford cables with a width of 3 mm and a trapezoidal angle of 1.9 degrees are twisted of 0.5 mm NbTi strands. The cross section of the $\cos 2\theta$ quadrupole magnet is shown in Fig. 10. Each pole has 21 turns, and the design value of the operating current is 2065 A. With a reference radius of 9.8 mm, the high-order field harmonics are all below $3 \times 10^{-4}$ [21]. The comparison of main design parameters using $\cos 2\theta$ coils and racetrack coils is shown in TABLE VI.

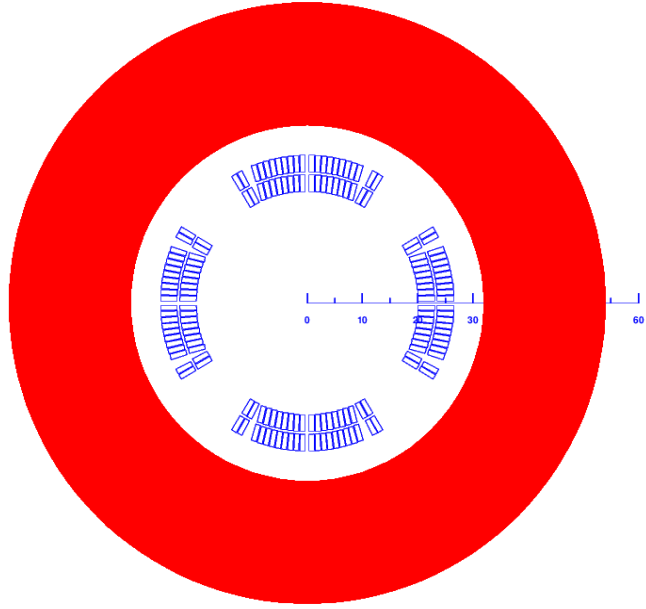

Fig. 10. Superconducting quadrupole magnet based on $\cos 2\theta$ coils.

TABLE VI
COMPARISON BETWEEN $\cos 2\theta$ AND RACETRACK QUADRUPOLE COILS

| Parameters | $\cos 2\theta$ (NbTi) | Racetrack (YBCO) |
|---|---|---|
| Current in cable (A) | 2065 | 745 |
| Current density in strands (A/mm$^2$) | 775 | 1043.692 |
| Ampere-turns per pole | 43365 | 52150 |
| Cross-sectional area of one coil (mm$^2$) | 58.59 | 36.4 |
| 2D Peak field in coil (T) | 3.192 | 4.574 |
| $B_{max}/(G*R)$ | 1.1708 | 1.6786 |

The comparison of the two coil structures shows that the $\cos 2\theta$ coil structure has higher excitation efficiency and lower peak field. In the case of using NbTi cable, the $\cos 2\theta$ coil structure has more advantages than racetrack magnets. But in the application of high-temperature superconducting materials, the racetrack quadrupole coil has more advantages. Firstly, from the perspective of the magnet structure, the racetrack coil has a simpler geometric shape. Then, fragile high-temperature superconductors are more suitable for racetrack coils with simple structures. Finally, $\cos 2\theta$ coils consist of low-temperature superconducting wires is more advantageous at low magnetic field strength. With the increase of magnetic field strength, the current-carrying capacity of low-temperature superconducting wires will be greatly reduced. However, the current-carrying capacity of high-temperature superconducting wires decreases slowly with the increase of magnetic field strength [22]. With the development of superconducting magnet technology, coil-dominated quadrupole magnets with racetrack coils composed of high-temperature superconductor have more potential in terms of high field gradient.

## VI. CONCLUSION

Superconducting quadrupole magnets using high-temperature superconductors are a frontier research direction, and coil-dominated quadrupole magnet based on racetrack coils is a feasible method to achieve a high gradient magnetic field. The theoretical formulas of each order field harmonics in coil-dominated racetrack quadrupole magnet are obtained, which are only related to the geometry of the racetrack coils. Using these expressions, we successfully find the solution of coil geometry parameters with the systematic field harmonics on the order of $10^{-4}$. The results of analytical method and FEM are consistent with each other. In application examples, we have completed the design of racetrack quadrupole magnet by using high-temperature superconductor YBCO.

Compared with the $\cos 2\theta$ magnet, the racetrack quadrupole magnet has a higher peak field and lower excitation efficiency. However, the racetrack quadrupole coil structure is more suitable for the strain-sensitive HTS superconductor. This study is beneficial for the application of coil-dominated superconducting quadrupole magnets based on racetrack coils in the future.

## APPENDIX A: EQUATION FOR FIELD GRADIENT AND HARMONICS

According to equation (17), the high-order multipole field $A_6$, $A_{10}$ and $A_{14}$ can be obtained.

$$A_2 = \frac{2R_{ref} u_0 j}{\pi} \ln \frac{(b^2+c^2)(a^2+d^2)}{(a^2+c^2)(b^2+d^2)}$$

$$A_6 = \frac{4R_{ref}^5 u_0 j}{5\pi}\left(\frac{6a^4+4a^2c^2+c^4}{24(a^2+c^2)^4}-\frac{6b^4+4b^2c^2+c^4}{24(b^2+c^2)^4}\right) - \frac{4R_{ref}^5 u_0 j}{5\pi}\left(\frac{6a^4+4a^2d^2+d^4}{24(a^2+d^2)^4}-\frac{6b^4+4b^2d^2+d^4}{24(b^2+d^2)^4}\right) - \frac{8R_{ref}^5 c^2 u_0 j}{\pi}\left(\frac{4a^2+c^2}{24(a^2+c^2)^4}-\frac{4b^2+c^2}{24(b^2+c^2)^4}\right) + \frac{8R_{ref}^5 d^2 u_0 j}{\pi}\left(\frac{4a^2+d^2}{24(a^2+d^2)^4}-\frac{4b^2+d^2}{24(b^2+d^2)^4}\right)$$
$$+ \frac{4R_{ref}^5 c^4 u_0 j}{\pi}\left(\frac{1}{8(a^2+c^2)^4}-\frac{1}{8(b^2+c^2)^4}\right) - \frac{4R_{ref}^5 d^4 u_0 j}{\pi}\left(\frac{1}{8(a^2+d^2)^4}-\frac{1}{8(b^2+d^2)^4}\right)$$

$$A_{10} = \frac{4R_{ref}^9 u_0 j}{9\pi}\left(\frac{70a^8+56a^6c^2+28a^4c^4+8a^2c^6+c^8}{560(a^2+c^2)^8}-\frac{70b^8+56b^6c^2+28b^4c^4+8b^2c^6+c^8}{560(b^2+c^2)^8}\right) - \frac{4R_{ref}^9 u_0 j}{9\pi}\left(\frac{70a^8+56a^6d^2+28a^4d^4+8a^2d^6+d^8}{560(a^2+d^2)^8}-\frac{70b^8+56b^6d^2+28b^4d^4+8b^2d^6+d^8}{560(b^2+d^2)^8}\right)$$
$$-\frac{112R_{ref}^9 c^6 u_0 j}{3\pi}\left(\frac{8a^2+c^2}{112(a^2+c^2)^8}-\frac{8b^2+c^2}{112(b^2+c^2)^8}\right) + \frac{112R_{ref}^9 d^6 u_0 j}{3\pi}\left(\frac{8a^2+d^2}{112(a^2+d^2)^8}-\frac{8b^2+d^2}{112(b^2+d^2)^8}\right) + \frac{4R_{ref}^9 c^8 u_0 j}{\pi}\left(\frac{1}{16(a^2+c^2)^8}-\frac{1}{16(b^2+c^2)^8}\right) - \frac{4R_{ref}^9 d^8 u_0 j}{\pi}\left(\frac{1}{16(a^2+d^2)^8}-\frac{1}{16(b^2+d^2)^8}\right)$$
$$+\frac{16R_{ref}^9 d^2 u_0 j}{\pi}\left(\frac{56a^6+28a^4d^2+8a^2d^4+d^6}{560(a^2+d^2)^8}-\frac{56b^6+28b^4d^2+8b^2d^4+d^6}{560(b^2+d^2)^8}\right) - \frac{16R_{ref}^9 c^2 u_0 j}{\pi}\left(\frac{56a^6+28a^4c^2+8a^2c^4+c^6}{560(a^2+c^2)^8}-\frac{56b^6+28b^4c^2+8b^2c^4+c^6}{560(b^2+c^2)^8}\right)$$
$$+\frac{56R_{ref}^9 c^4 u_0 j}{\pi}\left(\frac{28a^4+8a^2c^2+c^4}{336(a^2+c^2)^8}-\frac{28b^4+8b^2c^2+c^4}{336(b^2+c^2)^8}\right) - \frac{56R_{ref}^9 d^4 u_0 j}{\pi}\left(\frac{28a^4+8a^2d^2+d^4}{336(a^2+d^2)^8}-\frac{28b^4+8b^2d^2+d^4}{336(b^2+d^2)^8}\right)$$

$$A_{14} = \frac{4R_{ref}^{13} u_0 j}{13\pi}\left(\frac{924a^{12}+792a^{10}c^2+495a^8c^4+220a^6c^6+66a^4c^8+12a^2c^{10}+c^{12}}{11088(a^2+c^2)^{12}}-\frac{924b^{12}+792b^{10}c^2+495b^8c^4+220b^6c^6+66b^4c^8+12b^2c^{10}+c^{12}}{11088(b^2+c^2)^{12}}\right)$$
$$-\frac{4R_{ref}^{13} u_0 j}{13\pi}\left(\frac{924a^{12}+792a^{10}d^2+495a^8d^4+220a^6d^6+66a^4d^8+12a^2d^{10}+d^{12}}{11088(a^2+d^2)^{12}}-\frac{924b^{12}+792b^{10}d^2+495b^8d^4+220b^6d^6+66b^4d^8+12b^2d^{10}+d^{12}}{11088(b^2+d^2)^{12}}\right)$$
$$-\frac{88R_{ref}^{13} c^{10} u_0 j}{\pi}\left(\frac{12a^2+c^2}{264(a^2+c^2)^{12}}-\frac{12b^2+c^2}{264(b^2+c^2)^{12}}\right) + \frac{88R_{ref}^{13} d^{10} u_0 j}{\pi}\left(\frac{12a^2+d^2}{264(a^2+d^2)^{12}}-\frac{12b^2+d^2}{264(b^2+d^2)^{12}}\right)$$
$$+\frac{24R_{ref}^{13} d^2 u_0 j}{\pi}\left(\frac{792a^{10}+495a^8d^2+220a^6d^4+66a^4d^6+12a^2d^8+d^{10}}{11088(a^2+d^2)^{12}}-\frac{792b^{10}+495b^8d^2+220b^6d^4+66b^4d^6+12b^2d^8+d^{10}}{11088(b^2+d^2)^{12}}\right)$$
$$-\frac{24R_{ref}^{13} c^2 u_0 j}{\pi}\left(\frac{792a^{10}+495a^8c^2+220a^6c^4+66a^4c^6+12a^2c^8+c^{10}}{11088(a^2+c^2)^{12}}-\frac{792b^{10}+495b^8c^2+220b^6c^4+66b^4c^6+12b^2c^8+c^{10}}{11088(b^2+c^2)^{12}}\right)$$
$$+\frac{220R_{ref}^{13} c^4 u_0 j}{\pi}\left(\frac{495a^8+220a^6c^2+66a^4c^4+12a^2c^6+c^8}{7920(a^2+c^2)^{12}}-\frac{495b^8+220b^6c^2+66b^4c^4+12b^2c^6+c^8}{7920(b^2+c^2)^{12}}\right)$$
$$-\frac{220R_{ref}^{13} d^4 u_0 j}{\pi}\left(\frac{495a^8+220a^6d^2+66a^4d^4+12a^2d^6+d^8}{7920(a^2+d^2)^{12}}-\frac{495b^8+220b^6d^2+66b^4d^4+12b^2d^6+d^8}{7920(b^2+d^2)^{12}}\right) + \frac{4R_{ref}^{13} c^{12} u_0 j}{\pi}\left(\frac{1}{24(a^2+c^2)^{12}}-\frac{1}{24(b^2+c^2)^{12}}\right) - \frac{4R_{ref}^{13} d^{12} u_0 j}{\pi}\left(\frac{1}{24(a^2+d^2)^{12}}-\frac{1}{24(b^2+d^2)^{12}}\right)$$
$$-\frac{528R_{ref}^{13} c^6 u_0 j}{\pi}\left(\frac{220a^6+66a^4c^2+12a^2c^4+c^6}{3960(a^2+c^2)^{12}}-\frac{220b^6+66b^4c^2+12b^2c^4+c^6}{3960(b^2+c^2)^{12}}\right) + \frac{528R_{ref}^{13} d^6 u_0 j}{\pi}\left(\frac{220a^6+66a^4d^2+12a^2d^4+d^6}{3960(a^2+d^2)^{12}}-\frac{220b^6+66b^4d^2+12b^2d^4+d^6}{3960(b^2+d^2)^{12}}\right)$$
$$+\frac{396R_{ref}^{13} c^8 u_0 j}{\pi}\left(\frac{66a^4+12a^2c^2+c^4}{1320(a^2+c^2)^{12}}-\frac{66b^4+12b^2c^2+c^4}{1320(b^2+c^2)^{12}}\right) - \frac{396R_{ref}^{13} d^8 u_0 j}{\pi}\left(\frac{66a^4+12a^2d^2+d^4}{1320(a^2+d^2)^{12}}-\frac{66b^4+12b^2d^2+d^4}{1320(b^2+d^2)^{12}}\right)$$

Using the formula $a_n = A_n/A_2$, we can get systematic harmonics of the racetrack quadrupole coils.

$$a_6 = \frac{\begin{bmatrix}\frac{2R_{ref}^4}{5}\left(\frac{6a^4+4a^2c^2+c^4}{24(a^2+c^2)^4}-\frac{6b^4+4b^2c^2+c^4}{24(b^2+c^2)^4}\right) - \frac{2R_{ref}^4}{5}\left(\frac{6a^4+4a^2d^2+d^4}{24(a^2+d^2)^4}-\frac{6b^4+4b^2d^2+d^4}{24(b^2+d^2)^4}\right) - 4R_{ref}^4 c^2\left(\frac{4a^2+c^2}{24(a^2+c^2)^4}-\frac{4b^2+c^2}{24(b^2+c^2)^4}\right) + 4R_{ref}^4 d^2\left(\frac{4a^2+d^2}{24(a^2+d^2)^4}-\frac{4b^2+d^2}{24(b^2+d^2)^4}\right) \\ +2R_{ref}^4 c^4\left(\frac{1}{8(a^2+c^2)^4}-\frac{1}{8(b^2+c^2)^4}\right) - 2R_{ref}^4 d^4\left(\frac{1}{8(a^2+d^2)^4}-\frac{1}{8(b^2+d^2)^4}\right)\end{bmatrix}}{\ln\frac{(b^2+c^2)(a^2+d^2)}{(a^2+c^2)(b^2+d^2)}}$$

$$a_{10} = \frac{\begin{bmatrix} \frac{2R_{ref}^8}{9}\left(\frac{70a^8+56a^6c^2+28a^4c^4+8a^2c^6+c^8}{560(a^2+c^2)^8}-\frac{70b^8+56b^6c^2+28b^4c^4+8b^2c^6+c^8}{560(b^2+c^2)^8}\right)-\frac{2R_{ref}^8}{9}\left(\frac{70a^8+56a^6d^2+28a^4d^4+8a^2d^6+d^8}{560(a^2+d^2)^8}-\frac{70b^8+56b^6d^2+28b^4d^4+8b^2d^6+d^8}{560(b^2+d^2)^8}\right) \\ -\frac{56R_{ref}^8c^6}{3}\left(\frac{8a^2+c^2}{112(a^2+c^2)^8}-\frac{8b^2+c^2}{112(b^2+c^2)^8}\right)+\frac{56R_{ref}^8d^6}{3}\left(\frac{8a^2+d^2}{112(a^2+d^2)^8}-\frac{8b^2+d^2}{112(b^2+d^2)^8}\right)+2R_{ref}^8c^8\left(\frac{1}{16(a^2+c^2)^8}-\frac{1}{16(b^2+c^2)^8}\right)-2R_{ref}^8d^8\left(\frac{1}{16(a^2+d^2)^8}-\frac{1}{16(b^2+d^2)^8}\right) \\ +8R_{ref}^8d^2\left(\frac{56a^6+28a^4d^2+8a^2d^4+d^6}{560(a^2+d^2)^8}-\frac{56b^6+28b^4d^2+8b^2d^4+d^6}{560(b^2+d^2)^8}\right)-8R_{ref}^8c^2\left(\frac{56a^6+28a^4c^2+8a^2c^4+c^6}{560(a^2+c^2)^8}-\frac{56b^6+28b^4c^2+8b^2c^4+c^6}{560(b^2+c^2)^8}\right) \\ +28R_{ref}^8c^4\left(\frac{28a^4+8a^2c^2+c^4}{336(a^2+c^2)^8}-\frac{28b^4+8b^2c^2+c^4}{336(b^2+c^2)^8}\right)-28R_{ref}^8d^4\left(\frac{28a^4+8a^2d^2+d^4}{336(a^2+d^2)^8}-\frac{28b^4+8b^2d^2+d^4}{336(b^2+d^2)^8}\right) \end{bmatrix}}{\ln\frac{(b^2+c^2)(a^2+d^2)}{(a^2+c^2)(b^2+d^2)}}$$

$$a_{14} = \frac{\begin{bmatrix} \frac{2R_{ref}^{12}}{13}\left(\frac{924a^{12}+792a^{10}c^2+495a^8c^4+220a^6c^6+66a^4c^8+12a^2c^{10}+c^{12}}{11088(a^2+c^2)^{12}}-\frac{924b^{12}+792b^{10}c^2+495b^8c^4+220b^6c^6+66b^4c^8+12b^2c^{10}+c^{12}}{11088(b^2+c^2)^{12}}\right) \\ -\frac{2R_{ref}^{12}}{13}\left(\frac{924a^{12}+792a^{10}d^2+495a^8d^4+220a^6d^6+66a^4d^8+12a^2d^{10}+d^{12}}{11088(a^2+d^2)^{12}}-\frac{924b^{12}+792b^{10}d^2+495b^8d^4+220b^6d^6+66b^4d^8+12b^2d^{10}+d^{12}}{11088(b^2+d^2)^{12}}\right) \\ -44R_{ref}^{12}c^{10}\left(\frac{12a^2+c^2}{264(a^2+c^2)^{12}}-\frac{12b^2+c^2}{264(b^2+c^2)^{12}}\right)+44R_{ref}^{12}d^{10}\left(\frac{12a^2+d^2}{264(a^2+d^2)^{12}}-\frac{12b^2+d^2}{264(b^2+d^2)^{12}}\right) \\ +12R_{ref}^{12}d^2\left(\frac{792a^{10}+495a^8d^2+220a^6d^4+66a^4d^6+12a^2d^8+d^{10}}{11088(a^2+d^2)^{12}}-\frac{792b^{10}+495b^8d^2+220b^6d^4+66b^4d^6+12b^2d^8+d^{10}}{11088(b^2+d^2)^{12}}\right) \\ -12R_{ref}^{12}c^2\left(\frac{792a^{10}+495a^8c^2+220a^6c^4+66a^4c^6+12a^2c^8+c^{10}}{11088(a^2+c^2)^{12}}-\frac{792b^{10}+495b^8c^2+220b^6c^4+66b^4c^6+12b^2c^8+c^{10}}{11088(b^2+c^2)^{12}}\right) \\ +110R_{ref}^{12}c^4\left(\frac{495a^8+220a^6c^2+66a^4c^4+12a^2c^6+c^8}{7920(a^2+c^2)^{12}}-\frac{495b^8+220b^6c^2+66b^4c^4+12b^2c^6+c^8}{7920(b^2+c^2)^{12}}\right) \\ -110R_{ref}^{12}d^4\left(\frac{495a^8+220a^6d^2+66a^4d^4+12a^2d^6+d^8}{7920(a^2+d^2)^{12}}-\frac{495b^8+220b^6d^2+66b^4d^4+12b^2d^6+d^8}{7920(b^2+d^2)^{12}}\right) \\ +2R_{ref}^{12}c^{12}\left(\frac{1}{24(a^2+c^2)^{12}}-\frac{1}{24(b^2+c^2)^{12}}\right)-2R_{ref}^{12}d^{12}\left(\frac{1}{24(a^2+d^2)^{12}}-\frac{1}{24(b^2+d^2)^{12}}\right)-264R_{ref}^{12}c^6\left(\frac{220a^6+66a^4c^2+12a^2c^4+c^6}{3960(a^2+c^2)^{12}}-\frac{220b^6+66b^4c^2+12b^2c^4+c^6}{3960(b^2+c^2)^{12}}\right) \\ +264R_{ref}^{12}d^6\left(\frac{220a^6+66a^4d^2+12a^2d^4+d^6}{3960(a^2+d^2)^{12}}-\frac{220b^6+66b^4d^2+12b^2d^4+d^6}{3960(b^2+d^2)^{12}}\right) \\ +198R_{ref}^{12}c^8\left(\frac{66a^4+12a^2c^2+c^4}{1320(a^2+c^2)^{12}}-\frac{66b^4+12b^2c^2+c^4}{1320(b^2+c^2)^{12}}\right)-198R_{ref}^{12}d^8\left(\frac{66a^4+12a^2d^2+d^4}{1320(a^2+d^2)^{12}}-\frac{66b^4+12b^2d^2+d^4}{1320(b^2+d^2)^{12}}\right) \end{bmatrix}}{\ln\frac{(b^2+c^2)(a^2+d^2)}{(a^2+c^2)(b^2+d^2)}}$$